\documentclass[aps,showpacs,twocolumn,floats,prd,superscriptaddress,nofootinbib]{revtex4} 
\usepackage{graphicx,amsmath,amssymb,amstext}
\usepackage{amssymb,amsbsy,amsfonts,amsthm,color}

\usepackage{epsfig}
\usepackage{graphicx}
\usepackage{subfigure}

\graphicspath{{Figures/}}

\begin{document}

\title{Gravitational rotation of polarization: \\ 
Clarifying the gauge dependence and prediction for double pulsar}

\author{Ue-Li Pen}
\email{pen@cita.utoronto.ca}
\affiliation{Canadian Institute of Theoretical Astrophysics, 60 St George St, Toronto, ON M5S 3H8, Canada.}
\affiliation{Canadian Institute for Advanced Research, CIFAR program in Gravitation and Cosmology.}
\affiliation{Dunlap Institute for Astronomy \& Astrophysics, University of Toronto, AB 120-50 St. George Street, Toronto, ON M5S 3H4, Canada.}
\affiliation{Perimeter Institute of Theoretical Physics, 31 Caroline Street North, Waterloo, ON N2L 2Y5, Canada.}

\author{Xin Wang}
\email{xwang@cita.utoronto.ca}
\affiliation{Canadian Institute of Theoretical Astrophysics, 60 St George St, Toronto, ON M5S 3H8, Canada.}

\author{I-Sheng Yang}
\email{isheng.yang@gmail.com}
\affiliation{Canadian Institute of Theoretical Astrophysics, 60 St George St, Toronto, ON M5S 3H8, Canada.}
\affiliation{Perimeter Institute of Theoretical Physics, 31 Caroline Street North, Waterloo, ON N2L 2Y5, Canada.}

\begin{abstract}
From the basic concepts of general relativity, we investigate the rotation of the polarization angle by a moving gravitational lens. 
Particularly, we clarify the existing confusion in the literature by showing and explaining why such rotation must explicitly depend on the relative motion between the observer and the lens. 
We update the prediction of such effect on the double pulsar PSR J0737-3039 and estimate a rotation angle of $\sim 10^{-7}rad$. 
Despite its tiny signal, this is $10$ orders of magnitude larger than the previous prediction by Ruggiero and Tartaglia \cite{RugTar06}, which apparently was misguided by the confusion in the literature. 
\end{abstract}

\maketitle

\section{Introduction}

Einstein's theory of General Relativity has been the dominant theory of gravity for a century. 
Many of its signature outcomes, such as light-bending, orbital precession, and gravitational waves, have been confirmed by precision tests in astronomy \cite{DysEdd20,KraSta97,WeiTay04,Abb16}. 
One of the last pending tests is the gravitational rotation of the polarization angle. 
Since gravity directly affects the spacetime geometry, a gravitational lens not only can bend the light ray, it may also rotate the polarization.
Such gravitational rotation of polarization has not been measured so far.
\footnote{The E-mode and B-mode in CMB are defined as the relative angle between polarization and gradient. 
The observed leading order effect is a consequence of a rotated gradient but a fixed polarization, thus it does not count.
The actual rotation of polarization contributes to sub-leading effects, which are analyzed in \cite{Dai14}, and it is unclear whether we will eventually observe them.} 
One obvious reason is that it is usually very small. 
The rotation angle of a linearly polarized light ray, caused by a gravitational lens, is suppressed by two small numbers.
\begin{equation}
\Delta\phi \approx \frac{4GM}{r} \cdot v~.
\label{eq-v}
\end{equation}
Here $M$ is the mass of the lens, $r$ is the impact parameter---the shortest distance when the light ray pass near the lens, and $v$ is the velocity of the lens.  
Unless the light goes through somewhere comparable to the Schwarzschild radius, the first factor is small. 
Unless the velocity is almost relativistic, the second factor is small. 

Luckily, an almost edge-on, compact pulsar binary system, such as the double pulsar PSR J0737-3039, can be a very strong candidate to measure such effect. 
Pulsar signals are often highly polarized, allowing precise measurements of its rotation.
This system has an almost edge-on orbit which allows the impact parameter to be very small at the superior conjunction.
Finally, its compact orbit, with a two-hour period, means a large velocity. 
One main point of this paper is to show that one can expect to have $\Delta\phi\sim 10^{-7}$ from double pulsar, which might be observable given a dedicated observation campaign.

The gravitational rotation of polarization from this double pulsar system was previously studied in \cite{RugTar06}. 
They however derived a much smaller number which is incorrect. 
In fact, various theoretical derivations of this rotation of polarization have probably brought more confusion  
than clarity since its first appearance in \cite{Skr57}.
It was summarized in \cite{BroDem11} that three different values of $\Delta\phi$ can be derived from existing literature for seemingly identical physical situations.
Many authors disagreed on ``whether there is a nonzero rotation in Schwarzschild metric'', while none of them correctly pointed out that it is not even a well-defined question to ask.
Although Eq.~(\ref{eq-v}) has been derived by some authors, such as in \cite{KopMas01, Dai14}, they did not explicitly explain why it is the correct result.

In this paper, we will resolve the confusions by deriving Eq.~(\ref{eq-v}) from the very basic concept of general relativity---parallel transports. 
It turns out that $\Delta\phi$, despite being a number, is not a gauge-invariant scalar.
It is actually an $SO(2)$ projection of an $SO(3,1)$ tensor.
The $SO(3,1)$ rotation is a gauge-invariant property associated with a light ray that starts and ends far away from the lens, but $\Delta\phi$ depends on which $SO(2)$ we project to.
Therefore it is natural for $\Delta\phi$ to be gauge-dependent.
 {\it More precisely, the rotation of polarization depends on the frame of the observer. }
Such observer/gauge-dependence was the source of confusions.
For example, one cannot simply ask whether there are rotations in Schwarzschild metric without specifying who the observer is.
For an observer at rest, there is indeed zero rotation; 
for a moving observer, however, the rotation will be nonzero.

We also identify the ``correct'' gauge to compute such rotation, in which the answer will agree with the actual observation of double-pulsar signal from the earth.
It involves comparing the polarization of two signals, and both are measured in the rest frame of the earth.
Thus it is only natural that we calculate in the earth (observer's) frame.
The $v$ in Eq.~(\ref{eq-v}) is the relative velocity between the observer and the lens.

In Sec.\ref{sec-born}, we will provide an operational definition of polarization rotation from the basic principles in general relativity and explain the inevitable observer-dependence.
In Sec.\ref{sec-Sch}, we will derive Eq.~(\ref{eq-v}) in the small rotation limit for one spin-less, point-mass lens. 
We then generalize it into multiple lenses even with spins.
In Sec.\ref{sec-prediction}, we will describe the observable effects on double pulsar. 
In Sec.\ref{sec-obs}, we conclude with a discussion of an appropriate observation campaign of future detection.

\section{Definition from Scratch}
\label{sec-born}

\subsection{Parallel Transport around a Loop}

Intuitively, one can imagine the polarization as a vector attached to a light ray that is spacelike and orthogonal to the direction of propagation. 
Let {\bf k} be the null vector of the light ray and {\bf e} be a polarization vector, a parallel transport of {\bf e} should be valid in the geometric optic limit, i.e.
\begin{equation}
k^a \nabla_a e^c = k^a \partial_a e^c + k^a e^b \Gamma_{ab}^c =0~.
\label{eq-para}
\end{equation}
When the rotation is small, using the Born approximation, we can integrate along a light ray from point A to point B to get the change of polarization vector.
\begin{equation}
\Delta e^c = \int_A^B \hat{k}^a e_0^b \Gamma_{ab}^c~dl~,
\label{eq-int}
\end{equation}
where ${\bf e_0}$ is the original vector and $\Delta${\bf e} is the change. 
Straightforwardly, $\Delta \phi \equiv |\Delta {\bf e}|/|{\bf e}|$ could serve as a definition of how much a polarization vector has been rotated.
This however, cannot be the full story.
That is because Eq.~(\ref{eq-int}) literally compares two vectors on the tangent spaces of two different points,.
Such comparison is mathematically meaningless. 
The two vectors must be in the same tangent space to provide a meaningful rotation. 
Another way to state the same problem is that the value of connections, $\Gamma^c_{ab}$, depends on the coordinate choice.
By choosing a different gauge, one can change the value of the line integral of Eq.~(\ref{eq-int}) to any value.

Many literature used Eq.~(\ref{eq-int}) and computed its value in one very natural gauge, for example \cite{KopMas01, Dai14}.
We can probably call that the asymptotic Minkowski gauge or the Schwarzschild gauge, in which $\Gamma_{ab}^c$ falls of to zero asymptotically away from matter sources as they would in the Schwarzschild coordinate.
It turns out that Eq.~(\ref{eq-int}) in such gauge happens to give the correct, gauge-independent answer.
Here we will explain why.

First of all, parallel transport is not limited to null rays.
One vector $e^\mu$ at a point can be parallel transported along any path by an integral similar to Eq.~(\ref{eq-int}). 
After we parallel transport from $A$ to $B$ along a null ray, we can again parallel transport the resulting vector in $B$ along another curve back to $A$.
If we compare the final vector back at $A$ with the initial vector, the result is a loop integral.
\begin{equation}
\Delta e^c = \int_A^B \hat{k}^a e_0^b \Gamma_{ab}^c~dl +
 \int_B^A \hat{p}^a e_0^b \Gamma_{ab}^c dl~.
\label{eq-loop}
\end{equation}
Since now we are comparing two vectors on the tangent space of one point, it is mathematically meaningful.
It is also gauge invariant.
The gauge freedom allows us to change the value of $\Gamma_{ab}^c$ locally but not globally, as there will be constraints.
Parallel transport in any close loop is indeed one of those constraints.
Its answer carries the gauge-invariant information about spacetime curvature enclosed by such loop, and it must be gauge invariant.

Eq.~(\ref{eq-int}) is just providing a convenient way to evaluate such gauge-invariant loop integral.
Assume that we are in asymptotic Minkowski space, and both points $A$ and $B$ are in the asymptotic region.
We can then choose this path from $B$ back to $A$ to stay in the asymptotic region.
In the asymptotic Minkowski gauge that $\Gamma_{ab}^c\rightarrow0$ asymptotically, the segment of integral through the asymptotic region contributes nothing.
Therefore, the line integral, Eq.~(\ref{eq-int}), in the asymptotic Minkowski gauge, gives exactly the gauge-invariant answer of the loop integral in Eq.~(\ref{eq-loop}).

Furthermore, when there are multiple matter sources in asymptotic Minkowski space, there is a well-defined, common asymptotic Minkowski gauge.
That simply means $\Gamma_{ab}^c$ is only nonzero near sources, decays away from individual sources like in a Schwarzschild metric, and the contribution from each source superimposes in the region far away from all sources.
Computing a gauge-invariant loop integral in this gauge allows us to identify the contributions to rotation from individual sources, since only the segments of integral near sources have nonzero contributions.

\begin{figure}
\includegraphics[width=0.45\textwidth]{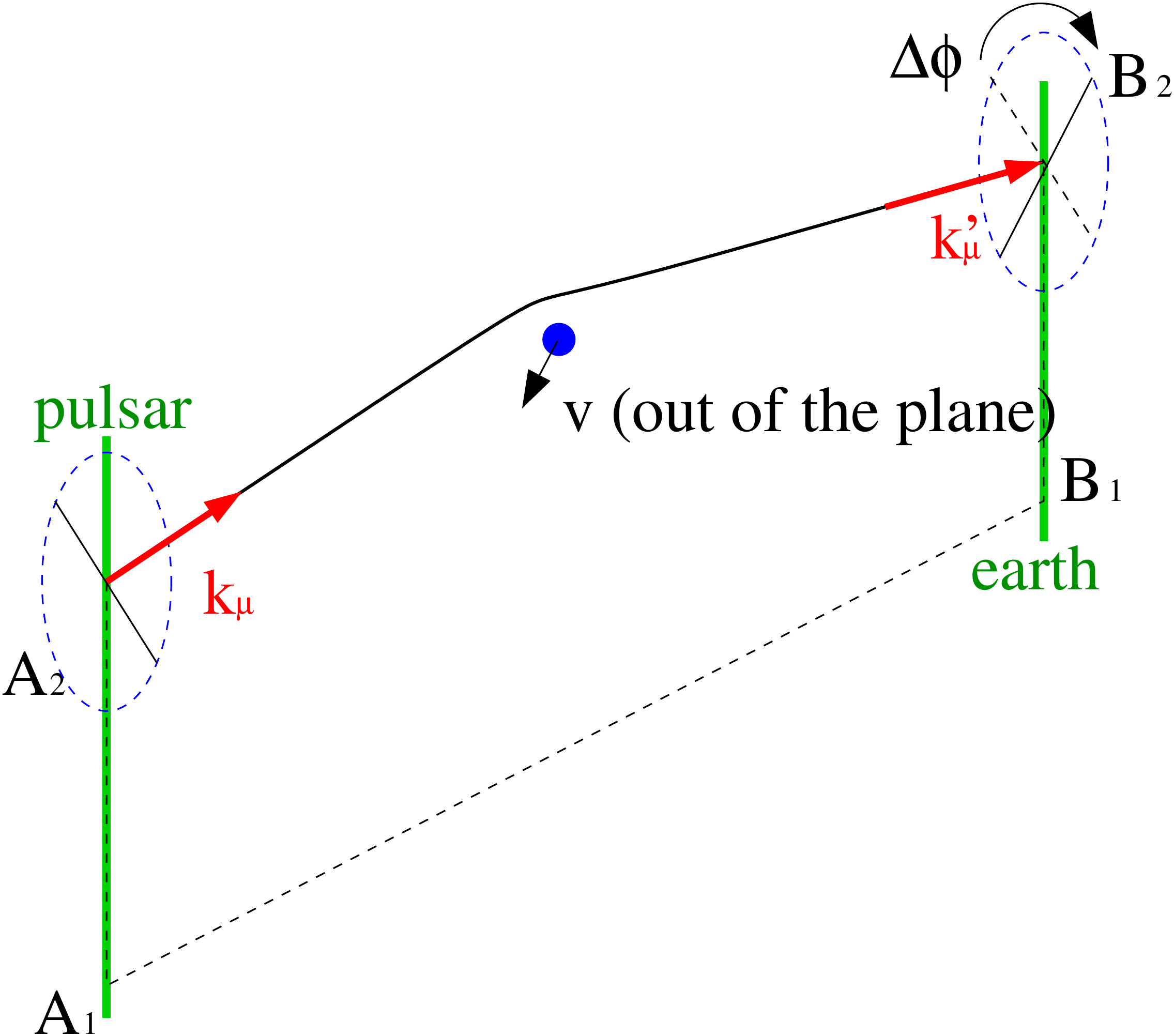}
\caption{\label{fig:loop}
Parallel transport along a loop contains 4 segments: bent light ray (solid black), worldline of the pulsar (thick, green, left), worldline of the observer (thick, green, right), and an unbent light ray (dashed, bottom). It leads to an $SO(3,1)$ rotation of the tangent space, and contains the information of both deflection of light (from $k_\mu$ to $k'_\mu$) and the rotation of polarization, $\Delta\phi$. In practice, we can measure this effect by comparing the pulsar signal when the companion neutron star (blue dot) passes through the line-of-site (solid line) to the same signal in other times (dotted line).
}
\end{figure}

An actual observation, as we depict that in Fig.\ref{fig:loop}, is closely related to a loop integral.
What we have is a source (pulsar) which constantly emits a fixed (albeit unknown) polarization. 
We measure the polarization during a usual time, which is a light ray from $A_1$ to $B_1$. 
And then we compare it with the polarization measured when its binary companion passes very close to the line of sight, which is another light ray from $A_2$ to $B_2$. 
We take the difference between these two measurements, which is a loop integral if we add two extra time-like segments.
\begin{eqnarray}
\Delta e^c &=& e^c|_{ B_2} - e^c|_{ B_1}
= \int_{A_2}^{B_2} \hat{k}^a e_0^b \Gamma_{ab}^c~dl
 + \int_{B_2}^{B_1} \hat{p}^a e_0^b \Gamma_{ab}^c dl  \nonumber \\ 
&&+ \int_{B_1}^{A_1} \hat{k}^a e_0^b \Gamma_{ab}^c~dl +
\int_{A_1}^{A_2} \hat{p}^a e_0^b \Gamma_{ab}^c dl~.
\label{eq-pulsar}
\end{eqnarray}
The two time-like segments, $A_1A_2$ and $B_1B_2$,  and the light ray $A_1B_1$, are all far away from the companion.
They all contribute nothing in the asymptotic Minkowski gauge.
Thus the above loop integral can be calculated in the asymptotic Minkowski gauge as only the line integral $A_2B_2$.
Such line integral actually only has contribution near the companion, thus it is indeed the rotation of polarization caused by the passage of the companion.
\footnote{
This simplified story is true when both the emitter and receiver are light so we can ignore their contribution to $\Gamma_{ab}^c$.
In reality, points $A1$, $A2$ are near the pulsar, and $B_1$, $B_2$ are on earth, so neither is in the asymptotic region.
Thus the time-like segments can be nonzero, and the null segments will have extra contribution near the end points.
Nevertheless, the extra contributions to the null segments will cancel each other.
The time-like contributions has nothing to do with the companion, and they are degenerate with an intrinsic variation of pulsar signal or the telescope receiving function.
We can simply observe and fit such behaviour when the companion is not passing through the line-of-sight, and subtract it from the data.
Thus treating them as in the asymptotic region is a simple way to show the contribution from the companion without loss of generality.}

\subsection{Observer Dependence}

In the previous section, we explained that a given vector, parallel-transported along a loop, goes through a gauge-invariant rotation.
In the asymptotic Minkowski gauge, such rotation can be calculated along a light ray.
This however, does not resolve all the confusion in the rotation of polarization.
The next problem is {\it which vector do we rotate?}
Polarization lives in a 2-dimensional plane, which cannot be uniquely determined by a light ray.
Thus there is no unambiguous answer to the question of ``how much rotation of polarization has a light ray gone through."
We need both the light ray and the observer's 4-velocity to determine the 2-dimensional plane on which the polarization lives.
Therefore, ``how much rotation'' must have an observer dependent answer.

Mathematically, we can see that Eq.~(\ref{eq-int}) is the leading order effect of a small rotation matrix.
\begin{equation}
e^c = e_0^c + \Delta e^c = \Lambda^c_{\ b} e_0^b = 
\left( g^c_{\ b} + \Delta^c_{\ b} \right) e_0^b~,
\end{equation}
where
\begin{equation}
\Delta_{cb} = \int_A^B \hat{k}^a \Gamma^d_{ab} g_{cd}dl~.
\end{equation}
By definition of a rotation matrix, $\Delta_{cb}$ has to be anti-symmetric, which can be verified explicitly.
\begin{eqnarray}
\Delta_{ac} = \int k^b\Gamma_{ab}^d g_{cd}d\lambda &=& 
\frac{1}{2} \int k^b \left(\partial_a g_{bc} + \partial_bg_{ac} - \partial_c g_{ab}\right)d\lambda
\label{eq-Delta} \nonumber \\
&=& \frac{1}{2} \int k^b \left(\partial_a g_{bc} - \partial_c g_{ab}\right)d\lambda~.
\end{eqnarray}
Note that we have to drop the boundary term in the above integral, which is allowed because this is effectively a loop integral as we explained in the previous section.
This demonstrates that there is actually a full $SO(3,1)$ rotation, $\Lambda^a_{\ b}\approx\left(g^a_{\ b} + \Delta^a_{\ b}\right)$, that is associated with a loop, therefore a light ray which starts and ends in asymptotic Minkowski region. 

This $SO(3,1)$ rotation is the gauge-invariant property associated with the light ray, but it does not yet determine the rotation of polarization, which is only an $SO(2)$.
It also contains extra information such as the deflection of the light ray itself. 
Particularly, one needs to specify a two-dimensional plane of polarization to determine which $SO(2)$ to project to. 
For any observer, the polarization vectors are orthogonal to both the incoming light ray and its own worldline. 
Thus a projection to the co-dimension-two surface orthogonal to the light ray and the observer 4-velocity is the desired $SO(2)$ rotation of polarization. 
{\it Therefore, it is natural and necessary that polarization rotation depends on the observer frame.} This also explains the $v$ dependence in Eq.~(\ref{eq-v}), which has to be the velocity of the lens in the observer's frame.

One last possible confusion is why such dependence is on the observer instead of the source, since they seem to play equivalent roles in the integral of Eq.~(\ref{eq-int}). 
Again, we remind the readers that the apparent line integral in Eq.~(\ref{eq-int}) is a convenient illusion.
The physically meaningful rotation is defined by a loop, where one compares a vector to its parallel transported outcome after going around the loop.
Thus there does exist one unique point at which the rotation is defined, which is where the two polarizations are being compared. 
In practice, we will have no idea about the actual polarization when the signal is emitted at the pulsar. 
All we know are the polarizations we received on earth, so this is the frame we have to choose.

\subsection{Beyond Born Approximation}

The actual effect we will calculate in the rest of this paper will be quite small, so Born approximation is justified. 
Nevertheless, the above abstract explanation must still be true beyond the Born approximation, and we will spend this subsection to demonstrate that. 

It is straightforward to actually solve the parallel transport equation, Eq.~(\ref{eq-para}), instead of using the Born-approximation integral in Eq.~(\ref{eq-int}). 
A loop-parallel transport back to the same point is obviously an $SO(3,1)$ rotation.
So one can see that up to getting the $SO(3,1)$ rotation, everything we said in the previous section directly generalizes beyond Born approximation.
The only question is that we have used the unique $k^\mu$ to determine the $SO(2)$ projection in the Born approximation.
Now the direction of light is also deflected significantly, $k^\mu \rightarrow k'^\mu$. 
Do we still have an unambiguous way to determine which $SO(2)$ to project into?

The answer is yes, and this is how we do it. 
First of all, the observer's 4-velocity reduces $SO(3,1)$ down to $SO(3)$.
The light rays, before and after the deflection, $k$ and $k'$, are also reduced down to two spacelike vectors in the observer's frame, $\kappa$ and $\kappa'$.
As long as $\kappa\neq-\kappa'$, there is a unique, minimal $SO(3)$ rotation that aligns them.
\footnote{Note that there are many rotations which can align them, but there is a unique minimal rotation, that is rotating along the direction orthogonal to both of them, $(\kappa\times\kappa')$.}
Aligning $\kappa$ and $\kappa'$ also merges their polarization planes, in which an $SO(2)$ rotation is uniquely defined. 
Thus, one can see that even beyond Born approximation, the rotation of polarization is still a well-defined, unambiguous, observer-dependent $SO(2)$ projection of a covariant $SO(3,1)$ tensor.

\section{Explicit calculation}
\label{sec-Sch}

\subsection{Point Mass}

We will treat the gravitational lens as a point mass and model it with a Schwarzschild metric in the isotropic form, expanded to the leading order of $(M/r)$. 
\begin{eqnarray}
g_{ab}dx^adx^b &=& -\left(1-\frac{2M}{r}\right)dt^2 \\ \nonumber
&+& \left(1+\frac{2M}{r}\right)  \left(dx^2+dy^2+dz^2\right), 
\label{eq-SchIso}
\end{eqnarray}
where  $r^2 = x^2 + y^2 + z^2$,  and the Newton constant $G$ is conveniently set to $1$.
Instead of studying an arbitrary light ray in the above coordinate, we will shift and boost this metric such that the lens has arbitrary position and velocity, and the relevant light ray is aways $x=t$.
In principle, we need six parameters, i.e. $(x_0,y_0,z_0,v_x,v_y,v_z)$. 
By applying symmetries, we can further simplify the problem so that eventually only three will be needed. 

First we use shift symmetries in $x$ and $t$ to set $x_0=0$, which simply means that $t=0$ is defined as the time when the light ray is closest to the lens. 
Next, we set $v_x=0$, so instead of letting the lens to have an $x$-velocity, the asymptotic observer who measures the polarization will have a nonzero $x$-velocity. 
This changes nothing because the light ray is in the $x$ direction, $k^\mu = (1,1,0,0)$.
Independent of what $x$-velocity the observer has, the plane orthogonal to both the light ray and the observer will be the $y$-$z$ plane.
Thus we are always calculating the rotation of polarization on the $y$-$z$ plane.
Finally, using rotational symmetry on the $y$-$z$ plan, we can set $v_z=0$, leaving the remaining three parameters to be $v_y=v$, $y_0$ and $z_0$. 

\begin{figure}
\includegraphics[width=0.45\textwidth]{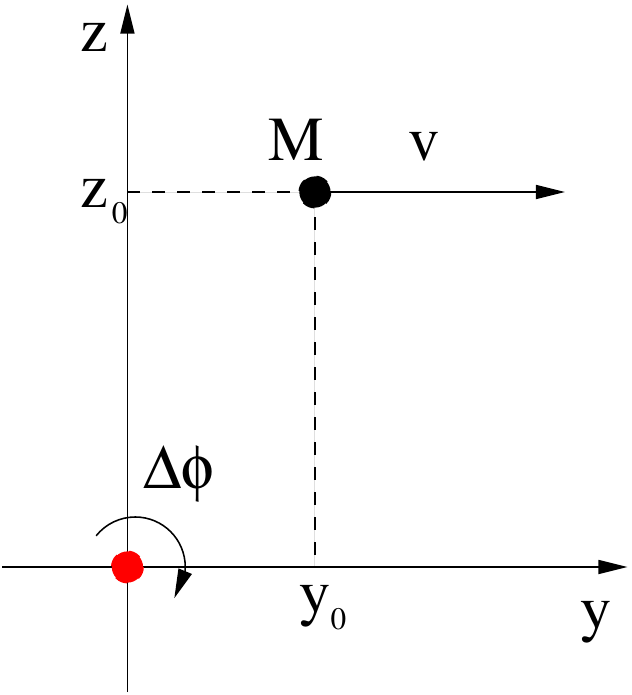}
\caption{\label{fig:rotation}
Using all symmetries, we can reduce any calculation of polarization-rotation from a spin-less, point-mass lens to this picture.
The red dot at the origin is the light ray traveling in the $x$ direction.
$M$ is the mass of the lens.
$v$ is its velocity (only in $y$ direction).
$(y_0,z_0)$ is the location of the lens relative to the light ray.
$\Delta\phi$ is the rotation of polarization, which is drawn in the appropriate direction in the picture.
One can visualize it as being ``dragged'' by the motion of the lens.
}
\end{figure}

Employing these symmetries significantly simplifies the problem, and we illustrate the final situation in Fig.\ref{fig:rotation}
Since the lens is the centre of the coordinate in Eq.~(\ref{eq-SchIso}), we need to apply the appropriate coordinate transformation to accommodate our symmetry choice. 
\begin{eqnarray}
\gamma = (1-v^2)^{-1/2}~, \ \ t \rightarrow \gamma(t-vy)~, \nonumber \\
 y\rightarrow \gamma(y-vt-y_0)~, \ \ z\rightarrow (z-z_0)~.
\end{eqnarray}
The resulting metric becomes
\begin{eqnarray}
\label{eq-metric}
g_{ab}dx^adx^b &=& -\left[1-\gamma^2(1+v^2)\frac{2M}{r}\right]dt^2   
\nonumber \\
&+& \biggl [1  +  \gamma^2(1+v^2)\frac{2M}{r} \biggr]dy^2 
- \frac{8Mv\gamma^2}{r}dtdy  \nonumber \\
&+& \left(1+\frac{2M}{r}\right)(dx^2+dz^2)~,
\end{eqnarray}
where $r = \sqrt{ \gamma^2(y-vt-y_0)^2 + x^2 + (z-z_0)^2 }$. 

While calculating the connections,
\begin{equation}
\Gamma_{ab}^c = \frac{g^{cd}}{2}\left(\partial_a g_{bd} + \partial_b g_{ad} - \partial_d g_{ab} \right)~,
\end{equation}
we can treat the first $g^{cd}$ as the flat metric $\eta^{cd}$ since we are only keeping the leading order result. 
This applies to any $g^{ab}$ that is not hit by a derivative in the calculation, e.g. the one in Eq.~(\ref{eq-Delta}).
We also assume that both the null ray direction and the polarization direction are only changed by a small amount, i.e. the Born approximation. 
Thus we can compute $\Delta_{ab}$ by Eq.~(\ref{eq-Delta}) along the undeflected light ray $x=t$.
Many components of $g_{ab}$ are zero due to our choice of symmetries, so it is straightforward to see that
\begin{eqnarray}
\Delta_{zy} &=& -\Delta_{yz} = \frac{1}{2}\int_{-\infty}^{\infty} dt \partial_z g_{ty} \nonumber \\
&=&-2Mv\gamma^2 z_0 \int_{-\infty}^{\infty}
\frac{dt}{\left[\gamma^2(vt+y_0)^2+t^2 + z_0^2\right]^{3/2}}. ~~~~
\end{eqnarray}

For $v\ll1$, we can perform the integral and keep only the leading order value to get 
\begin{equation}
\Delta\phi \equiv |\Delta_{yz}| \approx \frac{4Mvz_0}{y_0^2 + z_0^2}
= \frac{4|J_x|}{r_0^2}
= \frac{4 M |k\cdot(r_0\times v)|}{r_0^2}~.
\label{eq-gen}
\end{equation}
Here $J$ is the lens' angular momentum with respect to $x=0$, i.e. the point where the light ray was closest to the lens.
The direction of the light ray $k^a$ determines which component of $J$ we care about, which is the $x$-component in our symmetry choice.
The final, generalized format of Eq.~(\ref{eq-gen}) should be straightforward from our symmetry choice.
A more explicit calculation in \cite{KopMas01} led to the same result.

Furthermore, let us imagine that there is a continuous source of signals.
The light along the $x$ direction continues to shine while the lens is moving in its constant velocity in the $y$ direction.
The light ray which is closest to the lens will get the maximum rotation.
In other words, we get maximum rotation when the lens' velocity is orthogonal to its distance to the light way.
This corresponds to $y_0=0$ in the above calculation.
\begin{equation}
\Delta \phi_{Max} = \frac{4M}{z_0}v ~.
\end{equation}
This maximum rotation along a source trajectory is the promised result in Eq.~(\ref{eq-v}).

\subsection{General Case}

The above point-mass calculation assumes that it carries no spin. 
Many papers employed a Kerr metric instead to calculate how the angular momentum from the spin also contributes to the rotation of polarization. 
In the limit of small rotations, we can instead generalize the above result without explicitly starting from a Kerr metric. 
That is because Eq.~(\ref{eq-metric}) allows superposition when all lenses are not moving too fast and not too close to the light ray. 
At the leading order (ignoring sub-leading velocity terms), the metric of multiple moving point masses are given by
\begin{eqnarray}
g_{ab}dx^adx^b &=& -\left( 1 - 2\sum_n \frac{m^{(n)}}{r^{(n)}} \right)dt^2
 \\ \nonumber
&+& \left( 1 + 2\sum_n \frac{m^{(n)}}{r^{(n)}} \right) (dx^2+dy^2+dz^2) 
\\ \nonumber
&-& 8\sum_i \frac{m^{(n)} }{r^{(n)}}
\biggl(v^{(n)}_x dx  + v^{(n)}_y dy+ v^{(n)}_z dz\biggr)~, \\
r^{(n)} &=& \bigg[\left(x - x_0^{(n)} - v^{(n)}_x t\right)^2 
\\ \nonumber
& & + \left(y - y_0^{(n)} - v^{(n)}_y t\right)^2 
+ \left(z - z_0^{(n)} - v^{(n)}_z t\right) \bigg]^{1/2}
\end{eqnarray}

Their contributions to the total rotation also superimpose linearly. 
If we further assume that the masses are distributed in a small enough region such that their locations stay the same during the passage of the light ray, the answer is very simple. 
\begin{equation}
\Delta\phi = 4 \left| \sum_n m^{(n)} \frac{v_y^{(n)}z_0^{(n)} - v_z^{(n)}y_0^{(n)}}
{\left(y_0^{(n)}\right)^2+\left(z_0^{(n)}\right)^2} \right|~.
\label{eq-combine}
\end{equation}
Since their velocities are small, they are roughly in the same location after the light ray goes through all of them, thus $x_0^{(n)}$ do not matter at all. 

Under these assumptions, Eq.~(\ref{eq-combine}) provides the general answer to any mass and velocity distribution. 
By the uniqueness theorem, the effect from a Kerr metric of mass $M$ and spin $S$ can be mimicked by a two-particle system at the leading order.
\begin{eqnarray}
v_z^{(1)} &=& v_z^{(2)} = 0~,  \ \ \ v_y^{(1)} = v -\delta v~, \ \ \ v_y^{(2)} = v + \delta v~, \nonumber \\
y_0^{(1)} &=& y_0^{(2)} = y_0~, \ \ \ z_0^{(1)} = z_0-d~, \ \ \ z_0^{(2)} = z_0 + d~, \nonumber \\
m^{(1)} &=& m^{(2)} = M/2~, \ \ \ S = M (\delta v) d~. 
\end{eqnarray}
Taking $d\rightarrow0$ while holding $S$ fixed, we get
\begin{eqnarray}
\Delta \phi = 4\left|\frac{Mvz_0 + S}{y_0^2 + z_0^2}\right| = 4 \frac{\left| J_x + S_x \right|}{r_0^2}~.
\label{eq-spin}
\end{eqnarray}
The spin of the point mass contributes in exactly the same way as its ``orbital'' angular momentum around the light ray. Intriguingly, although \cite{KopMas01} agrees with Eq.~(\ref{eq-gen}), they claimed that the spin does not contribute at all. We cannot see any physical reason for such statement since Eq.~(\ref{eq-spin}) seems to be the most natural result.

\section{Example: Double Pulsar}
\label{sec-prediction}

\begin{figure}
\includegraphics[width=0.48\textwidth]{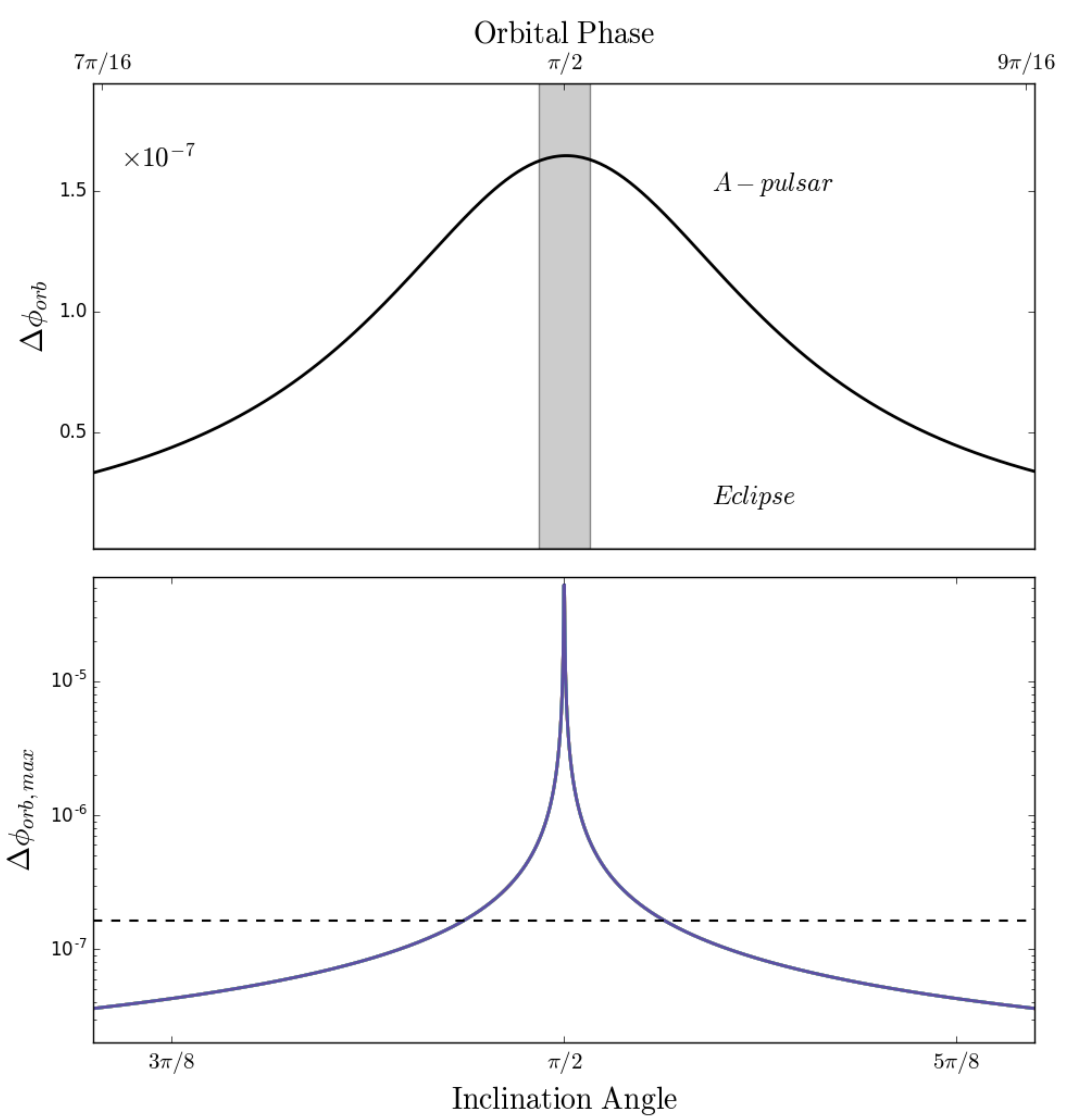}
\caption{\label{fig:rotang}
The angle of polarization rotation for binary pulsar PSR J0737-3039, the signals are shifted to peak at $\pi/2$ only for aesthetic reason. 
{\it Top:} Rotation angle $\Delta \phi$ of millisecond pulsar $A$, the grey area highlights the region blocked by eclipse. 
{\it Bottom:}  The inclination dependence on the peak rotation, assuming all other orbital parameters unchanged. The dashed line indicates the amplitude of real signal.
 }
\end{figure}

Before putting in the actual numbers, let us first give a rough estimation on the maximal rotation we can get from the double pulsar system.
Recall that the Schwarzschild radius of the sun is roughly $3km$, and these neutron stars are slightly larger.
A typical neutron star is slightly smaller than 10 times its own Schwarzschild radius, so we take the radius of the lens neutron star to be $30km$.
If it is a slow pulsar, we take the spin period to be about $1s$.
If the spin aligns with the line of sight, we estimate its contribution in Eq.~(\ref{eq-spin}) as
\begin{equation}
|S_x|  \sim  3km \times \frac{30km/1s}{c} \times 30km \sim 10^4 m^2~. \nonumber \\
{\color{red}   
}
\end{equation}
We have used both $c$ and $G$ to make this quantity to have the unit of length$^2$, which makes it easier to calculate the unitless $\Delta\phi$.
Similarly, assume the binary orbit is $10^9 m$, velocity is about $0.1\%$ speed of light, and the orbital tilt is $2$ degree, so the impact parameter is roughly $10^9 \times 2\pi/180 \approx 3.5 \times 10^7 m$, then the orbital contribution is roughly
\begin{equation}
|J_x| \approx  3km \times 10^{-3} \times 10^9 \times 2\pi/180  \approx 10^7 m^2~.
\end{equation}
In this case, the spin contribution is negligible.

If the lens is a recycled (fast) neutron star, its the period would be $\sim 1ms$, and its spin angular momentum is increased by a factor of $\sim1000$.
If it is also aligned with the line of sight, the spin contribution will be comparable to the orbital contribution.
Fortunately, fast pulsars are usually spun-up by accretion from the companion, so its spin is usually aligned with the orbital plane.
In our case, it means that the spin is almost perpendicular to the line-of-sight, so its contribution is again negligible.

Therefore, in either case, we can estimate the maximal rotation from the orbital contribution only, which is about\begin{equation}
\Delta\phi \approx 4 \frac{10^7 m^2}{\left[  10^9m\times  2\pi / 180 \right]^2}\approx 3*10^{-8}~.
\end{equation}
The actual value for double pulsar is slightly larger.
In Fig.~(\ref{fig:rotang}), we calculated the signal $\Delta \phi$ using the orbital information of double pulsar system PSR J0737-3039 \cite{KraSta97}. 
We ignore the spin contribution since they are negligible as we explained. 
We can see that during a rotation period, we can expect a maximal rotation of polarization ($\Delta\phi$) at about $10^{-7} rad$. 
This happens when the companion (lens) is almost in front of the pulsar.
It is well-known that at this moment, there will also be an eclipse, so one may worry that we cannot actually see this maximal rotation.
We specifically zoomed in and blacked-out the eclipsed.
We found that the peak of the $\Delta\phi$ curve is significantly wider than the eclipse duration.
Thus for a (relatively) long duration before and after the eclipse, we can still observe $\Delta\phi\sim10^{-7}rad$.
We can also see that although the double-pulsar is already quite edge-on, one can hope to get luckier and discover another system whose inclination angle is even closer to $90$ degree.
The resulting rotation of polarization can be even larger.



\section{Observational considerations}
\label{sec-obs}

Such a small swing in polarization angle will be challenging to detect.
For parameters of PSR J0737-3039 with a polarized fraction of about 50
\% \cite{2004ApJ...615L.137D}, raw thermal sensitivity requires a
signal-to-noise of at least $10^7$ in polarization.  
For a pulsar self-noise dominated telescope such as FAST \cite{FAST} or SKA \cite{SKA} with a band width of $\sim 1$ GHz, it requires $10^6$ seconds of on target on-pulse integration to achieve a $5-\sigma$ detection.
This is a substantial commitment of resources.  
On top of that, there are practicalities one must consider carefully.
First of all, any given object can be only seen for a limited about of time each day.
This observable duration for FAST could be short at the low declination of PSR J0737-3039, so SKA is likely the more suitable facility. 
At a duty cycle of about 20\%, this requires a few years of observations with the full phase 2 telescopes.  
During superior conjunction, the companion's magnetosphere partially absorbs the pulsar, and plasma faraday effects may also complicate the analysis.
\footnote{In principle, the plasma effect is frequency dependent, therefore distinguishable from the gravitational effect which is achromatic.} 

Despite the substantial efforts required, the pulsar is likely to be monitored extensively for other reasons.
Thus a detection may be eventually achieved over decades of SKA operations.  
Alternatively, other more optimal systems may be discovered, for example pulsar-black hole systems.  
Our estimates indicates that this effect is in principle observable in the foreseeable future.  
The gravitational Faraday effect can be added to the wish-list of pulsar tests of general relativity.

\newcommand{\apjl}{ApJ}      

\acknowledgments

We thank John Antoniadis and Michael Kramer for informative discussions.
We also thank the hospitality of Max-Planck Institute for Radio Astronomy during the Scintillometry Workshop.
This work is supported by the Canadian Government through the Canadian Institute for Advance Research and Industry Canada, and by Province of Ontario through the Ministry of Research and Innovation. 
The Dunlap Institute is funded through an endowment established by the David Dunlap family and the University of Toronto.
\bibliography{all_active}

\end{document}